\documentstyle[aps,epsfig]{revtex}
\def\be{\begin{equation}}
\def\ee{\end{equation}}
\begin{document}
 \title{On the reduced density matrix for a chain of free electrons}
 \author{
Ingo Peschel\\
{\small Fachbereich Physik, Freie Universit\"at Berlin,} \\
{\small Arnimallee 14, D-14195 Berlin, Germany}}
 \maketitle
 \begin{abstract}
 The properties of the reduced density
 matrix describing an interval of $N$ sites in an infinite chain
 of free electrons are investigated. A commuting operator is found for
 arbitrary filling and also for open chains. For a half filled periodic
 chain it is used to determine the eigenfunctions for the dominant eigenvalues 
 analytically in the continuum limit. Relations to the critical six-vertex 
 model are discussed.

\end{abstract}

\vspace{2cm}

 Reduced density matrices for some portion of a larger system 
play a central role in the density-matrix renormalization group (DMRG) method 
\cite{White92,White93,DMRG}. 
For this reason, they have been investigated
for a number of model systems in the last years 
\cite{Pescheletal99,Chung/Peschel00,Chung/Peschel01,Cheong/Henley02,Cheong/Henley03}.
It was found that for free electrons or bosons the reduced density matrices have
exponential form $exp(-{\cal H})$ and thus look like thermal operators.
It was also realized that one can base the 
determination of $\cal H$ on the one-particle correlation
functions in the state one is studying \cite{Cheong/Henley02,Peschel03,Vidal03,Latorre03}.
However, for critical lattice systems, which are particularly interesting in view of DMRG 
applications, the properties of $\cal H$ have only been studied numerically so far. 
For free electrons hopping on a chain one can see in this way 
the typical finite-size effects on the spectra and the concentration of the eigenfunctions 
near the boundaries. Nevertheless, a more explicit solution, which also makes contact with 
some results found in field theory \cite{Gaite03} in calculations of entanglement entropies, 
would be desirable. \\

  In the following we present such a solution by looking at the hopping model 
in a somewhat different way. We show that there exists a relatively
simple operator $\cal T$ which commutes with the reduced density matrix for the ground state 
and thus has the same eigenfunctions. This holds for arbitrary filling and also for a subsystem
at the end of a semi-infinite chain. The case of a half-filled periodic chain is then studied in detail.
Working with $\cal T$ and taking a proper continuum limit allows to determine the general 
character of the single-particle eigenstates. For the low-lying states it is found that
the eigenvalues of  $\cal H$ and $\cal T$ even coincide up
to a scale factor. We also discuss the connection with the critical six-vertex model.
This allows to derive the spectrum of $\cal H$ by a conformal mapping. It also
allows to view the density matrix as a particular transfer matrix and the
commutation relation as a special case of similar relations which occur in the
treatment of integrable two-dimensional models. \\

\section{Basic formulae}

We consider  a system of free fermions hopping between neighbouring 
sites of an infinite linear chain. The corresponding Hamiltonian reads 
 \be
  \hat{\cal{H}}=- \hat{t} \sum_{n} (c_n^{\dagger} c_{n+1} + c_{n+1}^{\dagger} c_{n})
  \label{eqn:hop}
 \ee
where 
$\hat t$ is the hopping matrix element and the 'hat' denotes quantities of the 
total system. The ground state $|0>$ is
a Slater determinant corresponding to a certain filling factor $\bar{n}$. As subsystem
we take $N$ consecutive sites, labelled by $i,j=1,2,..N$.  The
reduced density matrix $\rho$ has the general form 
\be
   \rho= {\cal K} \exp{({-\cal H} )}
   \label{eqn:rho2}
 \ee
where $\cal K$ is a normalization constant and
 \be
  {\cal H}= \sum_{i,j} H_{ij} c_i^{\dagger} c_j
  \label{eqn:H1}
 \ee
 The matrix $H_{ij} $ is completely determined by the one-particle correlation function
 of the total system
  \be
  \hat{C}_{mn} = <0|\; c_m^{\dagger} c_n\; |0>
  \label{eqn:cf}
 \ee
 which in our case is given by $\hat{C}_{mn}=C(m-n)$ where
 \be
  C(m) = \int_{-\pi}^{\pi} \frac {dq} {2\pi}\; n(q)\; e^{iqm}
  \label{eqn:cf1}
 \ee
 and $n(q)= 1$ for $|q|< \pi \bar{n}$ and zero otherwise. This leads to
 \be
  C(m)=  \frac{sin({\pi}\bar{n}\;m)} {{\pi}\;m}
  \label{eqn:cf2}
 \ee
 If $C$ is the $N \times N$ submatrix of $\hat{C}$ where 
 the sites are restricted to the subsystem, one has \cite{Cheong/Henley02,Peschel03}
 \be
    H = ln{\;[(1-C)/C]}
  \label{eqn:HC}
 \ee
 Thus $H$ and $C$ have common eigenfunctions and their eigenvalues
 $\varepsilon_k$ and $\zeta_k$ are related according to (\ref{eqn:HC}). 
 The $\zeta_k$ lie between 0 and 1 while the $\varepsilon_k$ vary between
 $ -\infty$ and $\infty$. For large $N$, the spectrum of $C$ must approach that 
 of $\hat C$ which is given by the step function $n(q)$. For this reason, most
 of the $\zeta_k$ lie exponentially close to 0 and 1 \cite{Latorre03,Cheong/Henley03} 
 and $C$ is not easy to handle numerically.

 \section{Commuting matrices}
 
 This difficulty can be circumvented by working with a matrix which commutes
 with $C$ (and thus $H$), but has a simpler spectrum and also a simpler form. The 
 existence of such a matrix is suggested by the structure of $H_{ij} $ obtained from numerical 
 calculations. One finds that $H_{ij} $ has approximately, but not exactly, tridiagonal form.
 A similar situation occurs for the transfer matrix of the two-dimensional Ising model, and 
 there a commuting tridiagonal matrix exists \cite{Suzuki71}. Trying therefore
 \be
   T =  
    \left(  \begin{array} {ccccc}
      d_1 & t_1   &  &   & \\  
       t_1 & d_2 & t_2  &  &\\
       & t_2 & d_3 & t_3  &  \\
        & & \ddots & \ddots & \\
        & & & t_{N-1} & d_N
  \end{array}  \right)   
  \label{eqn:tri}
 \ee 
 one finds indeed that $[C,T] = 0$, if the coefficients are chosen as
 \be
  t_i = \frac{i}{N} \left[1 - \frac{i}{N}\right],\;\;\;\;
  d_i = - cos(\pi\bar{n})\; \frac{2i-1}{N} \left[1 - \frac{2i-1}{2N}\right] \;.
  \label{eqn:coeff1} 
 \ee
 The mirror symmetry of the subsystem is reflected directly in these coefficients.
 Their linear increase near the boundaries is analogous to the situation found
 for corner transfer matrices \cite{Baxter82,Thacker86,TruongP88,Davies88}. 
 Physically, the matrix $T$ describes a hopping model with Hamiltonian 
 \be
  {\cal T} =  -\sum_{i=1}^{N-1}  t_i (c_i^{\dagger} c_{i+1} + c_{i+1}^{\dagger} c_{i}) -
     \sum_{i=1}^{N}   d_i c_i^{\dagger} c_{i} 
  \label{eqn:opt}
 \ee
 where both the hopping and the single-site energies increase towards the middle 
 of the subsystem. In the case of half filling ($\bar{n} =1/2$),
 the diagonal terms are zero and one has only hopping. It should be noted that $T$
 is only determined up to a constant factor, and we have chosen a normalization
 where the largest coefficients are of order one.
 
 A similar result holds if one considers a total system which is semi-infinite and
 chooses as subsystem the first $N$ sites next to the end. This geometry
 corresponds to the one in DMRG calculations with open boundaries. In this case, 
 one has to replace $C_{i,j}=C(i-j)$ by the expression $C^s_{i,j} =C(i-j)-C(i+j)$ in 
 all formulae. 
 Then one finds for half filling, that $[C^s,T^s] = 0$ if the $t_i$ are given by
 \be
   t_i = \frac{N+1+i}{4N} \left[1 - \frac{i}{N} \right].\;  
  \label{eqn:coeff2} 
 \ee
 The hopping varies again linearly near the interface with the rest of the
 system and is maximal at the open boundary. Basically, $T^s$ corresponds
 to the right half of the system described by $T$ which is easy to understand in terms 
 of the geometry. In the following we study $T$ for the case of a half-filled system.
 
 \section{Eigenvectors}
 
 We now use $T$ to determine the common eigenvectors of $C$, $H$ and $T$. 
 From numerical calculations one sees that for the eigenvalues $\lambda$ 
 of $T$ which are smallest in magnitude the eigenvectors $\varphi$ show rapid oscillations.
 It is convenient to take these out and also to separate the amplitudes at even and 
 odd sites by writing
 \be
  \Phi_j = (-1)^{j+1} \varphi_{2j-1}, \;\; \Psi_j = (-1)^{j+1} \varphi_{2j}
  \label{eqn:phipsi1}
 \ee
 This leads to the coupled equations
 \be
 -t_{2j-2} \Psi_{j-1} + t_{2j-1} \Psi_{j} = \lambda \Phi_j,\;\;\
  t_{2j-1} \Phi_{j} - t_{2j} \Phi_{j+1} = \lambda \Psi_j,
  \label{eqn:phipsi2}
 \ee
 If $N$ is odd, one eigenstate can be found exactly. Then $\lambda=0$
 is an eigenvalue and the equations decouple. One finds that $\Psi_j=0$,
 while $\Phi_j$ follows from the the recursion relation
 \be
  \Phi_{j+1} = \frac {t_{2j-1}}{t_{2j}}\; \Phi_j,
  \label{eqn:rec}
 \ee
 which can be solved explicitly. This "central state" (the $\lambda$ are distributed
 symmetrically with respect to zero) follows from the structure of the matrix
 and exists also for $T^s$. It has also been found in the study of Harper's 
 equation where a similar matrix (with sinusoidal elements) appears \cite{Hatsugai96}. To see
 the form of $\Phi$, one can consider large $N$ and take a continuum limit by
 writing $(2j-1)/N-1/2N= x, t_{2j-1}= t(x),
 \Phi_j= \Phi(x)$. Expanding the quantities, one then finds the differential equation
 \be
  \Phi'(x) = -\frac {1} {2} \frac {t'(x)} {t(x)} \;\Phi(x)
  \label{eqn:diff1}
 \ee
 with the solution
 \be
  \Phi(x) = \frac {c} {\sqrt{t(x)}} = \frac {c} {\sqrt{x(1-x)}}
  \label{eqn:phizero}
 \ee
 which reflects directly the form of the hopping amplitude $t(x)$. This state was studied 
 numerically in \cite{Cheong/Henley03}.\\
 The same continuum limit can also be used to determine the low-lying eigenstates in
 general. The equations (\ref{eqn:phipsi2}) then become
 \be
  (2t \frac {d} {dx} + t') \Psi(x) = \mu \Phi(x),\;\
  -(2t \frac {d} {dx} + t')\Phi(x) = \mu \Psi(x),
  \label{eqn:diff2}
 \ee
 where $\mu=\lambda N$ absorbs the $1/N$ dependence of the $t_j$. Combining the 
 equations and putting
 \be
  \Phi(x) = \frac {\chi(x)} {\sqrt{t(x)}}
  \label{eqn:xi}
 \ee
 one finds for $\chi$
 \be
  -4 t \frac {d} {dx} t \frac{d} {dx} \chi = \mu^2 \chi
  \label{eqn:diff3}
 \ee
 The substitution
 \be
  u = \frac {1} {2}\; ln\;(\frac{x}{1-x}),
  \label{eqn:u}
 \ee
 then leads to the oscillator equation for $\chi(u)$
 \be
  \frac {d^2 \chi} {du^2} + \mu^2 \chi = 0.
  \label{eqn:osc}
 \ee
 The general form of $\Phi$ is therefore
 \be
  \Phi(x) = \frac {c} {\sqrt{x(1-x)}} \; sin \left [\;  \frac {\mu} {2} \; ln\;(\frac {x} {1-x}) + \alpha
   \;\right ]
  \label{eqn:sol}
 \ee
 The function $\Psi(x)$ is obtained by replacing the sine with minus cosine according to
 the relations (\ref{eqn:diff2}).
 Such logarithmic oscillations of the eigenfunctions were also noted in \cite{Callan94} and had 
 been found earlier for corner
 transfer matrices \cite{Davies/Pearce90,P/Truong91,P/Wunderling92} where matrices 
 with linearly increasing elements are involved. For the Ising model, the square root appears there, 
 too, and reflects the surface exponent $x_s = 1/2$, while for the Gaussian model there is no 
 such prefactor.
 To determine the allowed values for $\mu$, one needs proper boundary conditions.
 From (\ref{eqn:phipsi2}) one has $t_1\Psi_1 = \lambda\Phi_1$ and $t_1\Phi_{N/2} = 
 \lambda\Psi_{N/2}$. These lead to the values $\alpha = \pm \pi/4$ for the phase and
 asymptotically one finds
 \be
   \mu_k = \pm \;\frac {\pi}{2\; lnN}\; (2k+1),\;\;\; k= 0,1,2, ....
   \label{eqn:conf}
 \ee
 i.e. equidistant levels scaling as $1/lnN$. For a bosonic field, such a spectrum was also 
 found in \cite{Callan94}. However, this behaviour can be seen only for very large $N$. 
 For moderate $N$ of the order 10-100 or even 1000, the dispersion relation $\mu_k$ vs. $k$ 
 still shows some 
 curvature \cite{Chung/Peschel01,Cheong/Henley03} and the eigenvalues are found to vary  
 as $1/(lnN+b_k)$ rather than as $1/lnN$. The constant $b_0$ for the lowest state, for example, 
 is approximately 2.6. Therefore one has
 considerable finite-size corrections to the asymptotic law. These features are also known 
 from studies of corner transfer matrices \cite{P/Truong87,Davies/Pearce90}.
 
 The trigonometric functions $\chi$ vanish near one end of the subsystem and are maximal 
 near the other one. For each eigenvalue, they are related by symmetry such that 
 $\Phi(x) = \pm \Psi(1-x)$.

 \section{Direct treatment of C}
 
 With the experience from the treatment of $T$ one can also obtain the eigenvectors directly
 from $C$. Since $C$ is not a sparse matrix, this leads to an integral equation. Working
 again with $\Phi$ and $\Psi$, one has instead of (\ref{eqn:phipsi2}) the equations
 \be
  \sum_j  {D(2i-2j+1)}  \Phi_j  = \bar{\zeta}\; \Psi_i,\;\; 
  -\sum_j  {D(2i-2j-1)}  \Psi_j  = \bar{\zeta}\; \Phi_i,\\
  \label{eqn:sum1}
 \ee
 where $D(l) = 1/(\pi \;l)$ is the denominator of $C(l)$ and $\bar{\zeta} = \zeta - C(0) = \zeta -1/2$. 
 Combining these, gives a single equation of the 
 form
 \be
  \sum_j K_{ij} \Phi_j = \bar{\zeta^2}\; \Phi_i
  \label{eqn:sum2}
 \ee
 Before taking the continuum limit, one has to single out the diagonal terms $K_{ii}$ which
 are positive, while the others are negative. Since 
 \be
  K_{ii} = \frac {1} {\pi^2} \sum_{l=1}^{N/2} \frac {1} {[2(l-i) + 1]^2}
  \label{eqn:sum3}
 \ee
 depends only weakly on $i$ and $N$, one approximates it by its limit for large  $i$ and $N$,
 $K_{ii} = K = 1/4$. The equation in the continuum limit then becomes
 \be
   \int_{0}^{1} dx' K(x,x')\; \Phi(x') = (\bar{\zeta^2}- \frac{1} {4} )\; \Phi(x)
   \label{eqn:int1}
 \ee
 where
 \be
   K(x,x') = - \frac {1} {2 \pi^2} \frac {1} {x-x'} \left[\;ln \frac {x} {1-x} - ln \frac {x'} {1-x'}\;\right].
   \label{eqn:kern}
 \ee
 Using the same substitutions as before, one obtains
 \be
  - \frac {1} {2 \pi^2} \int_{-\infty}^{\infty} du' \frac {(u-u')} {sh(u-u')}\; \chi(u') =
   (\bar{\zeta^2}- \frac{1} {4} )\; \chi(u)
   \label{eqn:int2}
  \ee
  Since one has a difference kernel now, the equation can be solved by Fourier transformation.
  Writing $\chi$ as a trigonometric function of argument $\mu u$ as in the solution of 
  (\ref{eqn:osc}), one finds for $\zeta$ 
  \be
    \zeta = \frac{1} {2} \left[ 1 + th\;(\frac {\pi \mu} {2}) \right]
    \label{eqn:zeta}
  \ee
  The eigenvalues $\varepsilon$ of $H$ then follow from $\varepsilon = ln((1-\zeta)/\zeta)$ , 
  which gives the simple result
  \be
   \varepsilon = \pi \mu
   \label{eqn:eps}
  \ee
  i.e. up to a factor of $\pi$ they are the same as the eigenvalues of $NT$. For the low-lying states
  and in the continuum limit, therefore, the matrices $H$ and $T$ are proportional to each other
  \be 
     H = \pi N T
   \label{eqn:HT}
  \ee
  This can be checked numerically, and one finds indeed, that $\varepsilon$ and $\lambda$ fulfill
  this relation with high accuracy. For $N=16$, for example, the relative deviations are of the order 
  of $10^{-3}$
  for the lowest states. In general, however, the relation (\ref{eqn:HT}) holds only approximately,
  and the deviations increase for the larger eigenvalues. The close relation of the two matrices
  can be understood from the structure of $H$ which, to a first approximation, has the same
  bidiagonal form as $T$ with very similar matrix elements.

  \section{Connection to two dimensions}

  Using the Jordan-Wigner transformation,
  $\hat{\cal H}$ can be expressed in terms of spin one-half operators and becomes the 
  Hamiltonian of an XX chain   
  \be
   \hat{\cal H} = - 2\;\hat t \; \sum_{n} (\sigma_n^x \sigma_{n+1}^x + \sigma_n^y \sigma_{n+1}^y)
   \label{eqn:xx1}
  \ee
  where $\sigma_n^x,\;\sigma_n^y$ are Pauli matrices. It is known that this operator commutes
  with the row transfer matrix $\hat{\cal V}$ of the six-vertex model with
   $\Delta = 0$ \cite{Sutherland70,Baxter82}. This model also corresponds to two 
  uncoupled Ising lattices at their critical temperature. Moreover, the ground 
  state of $\hat{\cal H}$ is the state with maximal eigenvalue for the transfer matrix. It follows that one 
  can interpret the reduced density matrix $\rho$ as the partition function of such a system, having 
  infinite size and a cut in it, along which the spins are kept as free variables 
  \cite{Pescheletal99,Nishino99}.
  In previous work on non-critical models, this cut was considered
  as half-infinite and one was lead directly to standard corner transfer matrices 
  \cite{Pescheletal99,Chung/Peschel00}. In the present case, the cut has a finite length $N$.\\

  This property of $\rho$ can be used to rederive the low-lying spectrum of $\cal H$. For this
  purpose, one considers the Riemann plane which is obtained by stacking such systems and
  joining different sides of the cut. If the cut is between $x=-a$ and $x=a$, this is the Riemann
  plane of the function $ln(z-a)/(z+a)$. For correlations between points on the cut, but in different
  sheets, $\rho$ plays the role of a transfer matrix, and the asymptotic behaviour of such correlation
  functions is determined by the largest eigenvalues of $\rho$. Since one is at criticality, one can use 
  the conformal transformation 
  \be
   w = ln \;\frac {z-a} {z+a}
   \label{eqn:conf1}
  \ee
  to map the complete Riemann plane to the full $w$ plane. If one further cuts out small circles with
  radius $b<<a$ around the endpoints $z= \pm a$ of the cut, the map is to a vertical strip 
  $-u_0 < u < u_0$ in the
  $w$ plane, where $w=u+iv$ and $u_0 = ln(2a/b)$. A vertical correlation function in the strip between 
  points lying $\Delta v =  2\pi$ apart correponds to correlations between equivalent points
  in consecutive Riemann sheets and involves one power of $\rho$ there. This allows to determine the
  spectrum of $\rho$ (or $\cal H$) from the correlations in the strip as in \cite{P/Truong87}. For 
  free boundary conditions along the small circles, this gives the positive $\varepsilon$ as
  \be
   \varepsilon_k = 2 \pi \frac {\pi } {L} (x_s + k) ,\;\; k= 0,1,2 ...
   \label{eqn:conf2}
  \ee
  where $L = 2 u_0$ is the width of the strip. 
  Inserting the value $x_s= 1/2$ for the surface exponent of the six-vertex model one then has
  \be
   \varepsilon_k = \frac {\pi^2} {2\;ln(2a/b)} \;(2k +1),\;\; k= 0,1,2 ...
   \label{eqn:conf3}
  \ee
  If one identifies $2a$ with $N$ and sets $b=1$, this is the continuum result (\ref{eqn:conf}),
  (\ref{eqn:eps}) found previously. By choosing  $b<1$, one could also obtain an
  additive correction to $ln N$.
  
  One should mention that a similar mapping was used in \cite{Holzhey94} to calculate
  entanglement entropies in the context of black holes. In this case, one needs $\rho$ to 
  calculate the
  entropy via $S = -tr(\rho\; ln \rho)$. The result was that asymptotically $S$ diverges as
  $S = (c/3)\;ln N$, where $c$ (equal to one here) is the conformal anomaly. This behaviour 
  was verified recently for various spin chains \cite{Vidal03,Latorre03,Jin03} and can 
  be understood in terms of the logarithmic dependence of $\varepsilon_k$ on the size $N$
  \cite{Callan94}.  A discretized spectrum as in (\ref{eqn:conf3}) was found 
  previously for finite-size corner transfer matrices. Then $2a$ corresponds to 
  the outer radius of the system and $b$ to the inner one \cite{P/Truong87}. \\
  
  The same approach can also be used to discuss $\rho$ for the case of a subsystem
  at the end of a half-infinite chain. If the end is at $x=0$, one is lead to a half-plane $x > 0$ with
  a cut from $x=0$ to $x=a$, i.e. from the left edge into the system. This maps to the left half of
  the previous strip under (\ref{eqn:conf1}). Therefore, one has to multiply $\varepsilon_k$ in
  (\ref{eqn:conf3}) by a factor of 2. Such an increase
  of the single-particle eigenvalues (corresponding to a faster drop of the spectrum of $\rho$
  itself) can also be seen in numerical calculations on the lattice
  and is responsible for the better performance of the DMRG with open boundaries in the
  present case of a critical system. In the entanglement entropy,
  this gives a factor of 1/2 which one can interpret as resulting from the reduced number
  of interfaces between the subsystem and the rest.

  \section{Discussion}
  
  We have considered the reduced density matrix for a chain of free electrons in its ground state
  and have found
  results in two directions. On the one hand, we could derive the structure of the low-lying 
  eigenfunctions which are most important in DMRG calculations and also in general. These 
  have their largest amplitudes near the interface(s) between the subsystem and its surrounding.
  Away from the interface, they decay with a power law involving the surface exponent
  $x_s = 1/2$ of the six-vertex model. Thus the influence of the environment propagates
  far into the interior, as expected for a critical system. The spectrum of $\cal H$ was also
  obtained and checked by a conformal mapping in the associated two-dimensional model.
  These results complement previous numerical studies \cite{Chung/Peschel01,Cheong/Henley03}. \\
  
  On a more general level we found the operator $\cal T$, Eqn. (\ref{eqn:opt}), which commutes 
  with the reduced density matrix $\rho$. We used it here as a tool for obtaining the 
  eigenfunctions of $\rho$, but it is interesting in itself. Written in terms of Pauli matrices,
  it has a form analogous to (\ref{eqn:xx1}), namely (for half filling)
  \be
    {\cal T} = -2\;\sum_{i=1}^{N-1} t_i \;(\sigma_i^x \sigma_{i+1}^x + \sigma_i^y \sigma_{i+1}^y)
   \label{eqn:xx2}
  \ee 
  and describes a spatially inhomogeneous XX chain of finite length. The commutation relation
  $[\rho, {\cal T} ] = 0$ is completely analogous to the relation $[\hat{\cal V},\hat{\cal H}] = 0$ 
  between the row transfer matrix $\hat {\cal V}$ of the six-vertex model
  and the Hamiltonian $\hat {\cal H}$ of the total chain. It would be interesting, and give further
  insight, to derive it in a different way. One should mention that reduced density matrices have 
  actually been used in studies of integrable models via vertex operators or via the algebraic
  Bethe ansatz. In particular, the six-vertex model and the equivalent XXZ spin chain were treated 
  in \cite{Jimbo96,Kitanine00}. This lead to exact, but rather complicated 
  expressions for correlation functions along the cut. By contrast, the commuting operator
  gives a relatively clear physical picture of the problem under consideration.

\end{document}